\documentclass{iaus}
\usepackage{graphics}

  \checkfont{eurm10}
  \iffontfound
    \IfFileExists{upmath.sty}
      {\typeout{^^JFound AMS Euler Roman fonts on the system,
                   using the 'upmath' package.^^J}%
       \usepackage{upmath}}
      {\typeout{^^JFound AMS Euler Roman fonts on the system, but you
                   dont seem to have the}%
       \typeout{'upmath' package installed. iaus.cls can take advantage
                 of these fonts,^^Jif you use 'upmath' package.^^J}%
      }
  \else
  \fi

  \checkfont{msam10}
  \iffontfound
    \IfFileExists{amssymb.sty}
      {\typeout{^^JFound AMS Symbol fonts on the system, using the
                'amssymb' package.^^J}%
       \usepackage{amssymb}%

      }{}
  \fi

  \IfFileExists{amsbsy.sty}
    {\typeout{^^JFound the 'amsbsy' package on the system, using it.^^J}%
     \usepackage{amsbsy}}
    {\providecommand\boldsymbol[1]{\mbox{\boldmath $##1$}}}

\newsavebox{\astrutbox}
\sbox{\astrutbox}{\rule[-5pt]{0pt}{20pt}}

\newcommand\etal{\mbox{\textit{et al.}}}

\def\gsim{ \lower .75ex \hbox{$\sim$} \llap{\raise .27ex \hbox{$>$}} }
\def\lsim{ \lower .75ex \hbox{$\sim$} \llap{\raise .27ex \hbox{$<$}} }
\def\Mo{{\rm M_\odot}}

\title[Outskirts of Galaxy Clusters: intense life in the suburbs]
      {On the age-radius relation and orbital history of cluster galaxies}

\author[Ben Moore {\it et al.\/}]%
{Ben Moore$^1$%
\thanks{Present address: Institute for Theoretical Physics,
University of Z\"urich, Winterthurerstr. 190, CH-8057 Z\"urich, Switzerland},
J\"urg Diemand $^1$\break
\and Joachim Stadel$^1$}

\affiliation{$^1$Institute for Theoretical Physics,
University of Z\"urich, Winterthurerstr. 190, CH-8057 Z\"urich, Switzerland. 
Email: moore@physik.unizh.ch}

\pubyear{2004}
\volume{195}
\pagerange{1--8}
\date{?? and in revised form ??}
\jname{Outskirts of Galaxy Clusters: intense life in the suburbs}
\editors{A. Diaferio, ed.}
\begin{document}

\maketitle 

\begin{abstract}

We explore the region of influence of a galaxy cluster using numerical simulations
of cold dark matter halos. Many of the observed galaxies in a cluster are
expected to be infalling for the first time. 
Half of the halos at
distances of one to two virial radii today have previously orbited
through the cluster, most of them have even passed through the dense inner regions of 
the cluster. Some halos at distances of up to three times the virial radius
have also passed through the cluster core. 
We do not find a significant 
correlation of 
 ``infall age'' versus present day position 
for substructures and the scatter at a given position is very large. This relation may
be much more significant if we could resolve the 
physically 
overmerged galaxies in the central
region.

\end{abstract}

\firstsection 
\section{Introduction}

Are the morphologies of galaxies imprinted during an early and rapid formation epoch 
or are they due to environmental processes that subsequently transform galaxies between 
morphological classes?
The gravitational and hydrodynamical mechanisms that could perform such
transformations were proposed in the 1970's, before the key 
observational evidence for environmental dependencies 
was provided 
- the 
morphology-density relation and the Butcher-Oemler effect.
Many recent numerical simulations support these theoretical expectations.
However, until we have self-consistent numerical simulations that can follow the
structural evolution of galaxies within a large computational volume, we 
must resort to semi-analytic treatments or to studying the evolution
of galaxies within idealised numerical calculations. 

In this paper we
study the orbits and infall history of substructure halos within
a cold dark matter galaxy cluster. When we observe a cluster today
we see a single frame 
of its 
entire cosmic evolution.
What we would
like to know is for a given galaxy at a given position, what is its
likely orbit? Is it infalling for the first time? What are the environments
that may have pre-processed the galaxy? If it has already passed pericenter
at what epoch did it enter a cluster-like environment? What was its impact
parameter and velocity with respect to the cluster center? Are clusters
built up in an ``onion shell'' scenario such that the observed galaxies
trace an age-radius relation?
 We shall use the 
largest and highest resolution calculations of cold dark matter
galaxy clusters to address some of these questions. With over 60 million
particles in the high resolution region, up
to 25 million particles within the virial radius 
and high force resolution we can resolve the orbital histories
of many thousands of substructures and halos both within the cluster
and in the suburbs.
Related studies have been carried out recently:
\cite{Balogh} followed 
{\it particle} orbits in N-body simulations while \cite[Mamon \etal\ (2004)]{Mamon}
used analytical calculations and the $z=0$ snapshots of simulations
to estimate rebound radii. Shortly after this meeting,
other groups (\cite{Gill}, 
\cite[Gao \etal\ (2004)]{Gao}) have published 
results that were also obtained by following 
{\it subhalo} orbits and their results are very similar to those
presented here.

\section{Accretion redshift of cluster subhalos}

It is interesting to know how much time todays cluster galaxies 
spent in dense environments and if the accretion time into a more massive halo 
is correlated with the current position in the cluster.
One could expect that subhalos which fell into the cluster 
(or one of it progenitors) early have less orbital energy and tend to end up 
closer to the cluster center.
We analyze the redshift of accretion of cluster subhalos in $\Lambda$CDM simulations.
Note that as accreted structures we count both subhalos of the final cluster and 
subhalos of the cluster progenitor groups. 

We take 20 outputs of run D6h and 10 of run C9, equally spaced in time.
The simulations are described in \cite{Diemand2004b} and the properties of
their subhalos are presented in \cite{Diemand2004a}. Run C9 resolves 
a $M_{\rm virial}= 5.0\times 10^{14}\Mo$ cluster with 10 million particles
within $r_{\rm virial}$ and D6h resolves a smaller
$M_{\rm virial}= 3.1\times 10^{14}\Mo$ cluster with 2 million particles.
 
\begin{figure}
\includegraphics{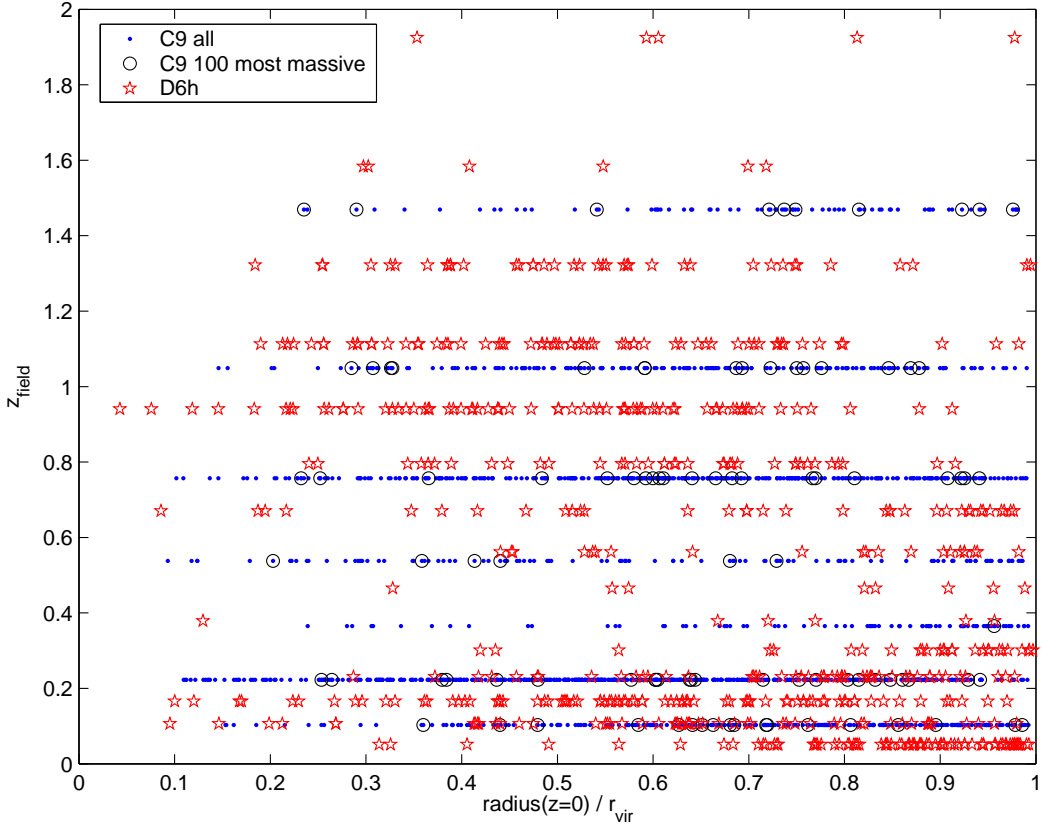}
\caption{
Accretion redshift of subhalos versus distance fromm the cluster 
center
today. We plot the redshift of the snapshot where a halo was identified as
an individual field halo for the last time. The trend that central 
subhalos spent more
time within the cluster is weak and the is a large scatter in accretion 
redshifts at all radii.}
\label{zaccretedscatter}
\end{figure}

The subhalos were identified with SKID (\cite[Stadel 2001]{Stadel2001})
and here we consider only structures with at least 32 bound particles. 
For each snapshot we construct a halo catalogue with FOF using a 
comoving linking length of 
0.164 $\Delta x_0$  and trace back in time all subhalos within the virial 
radius of todays cluster. In Figure \ref{zaccretedscatter} the redshift
before accretion is plotted, that is the last time a halo is identified
as individual field halo. There is a large scatter in the accretion
redshifts and no strong correlation with radius.
 
\begin{figure}
\includegraphics{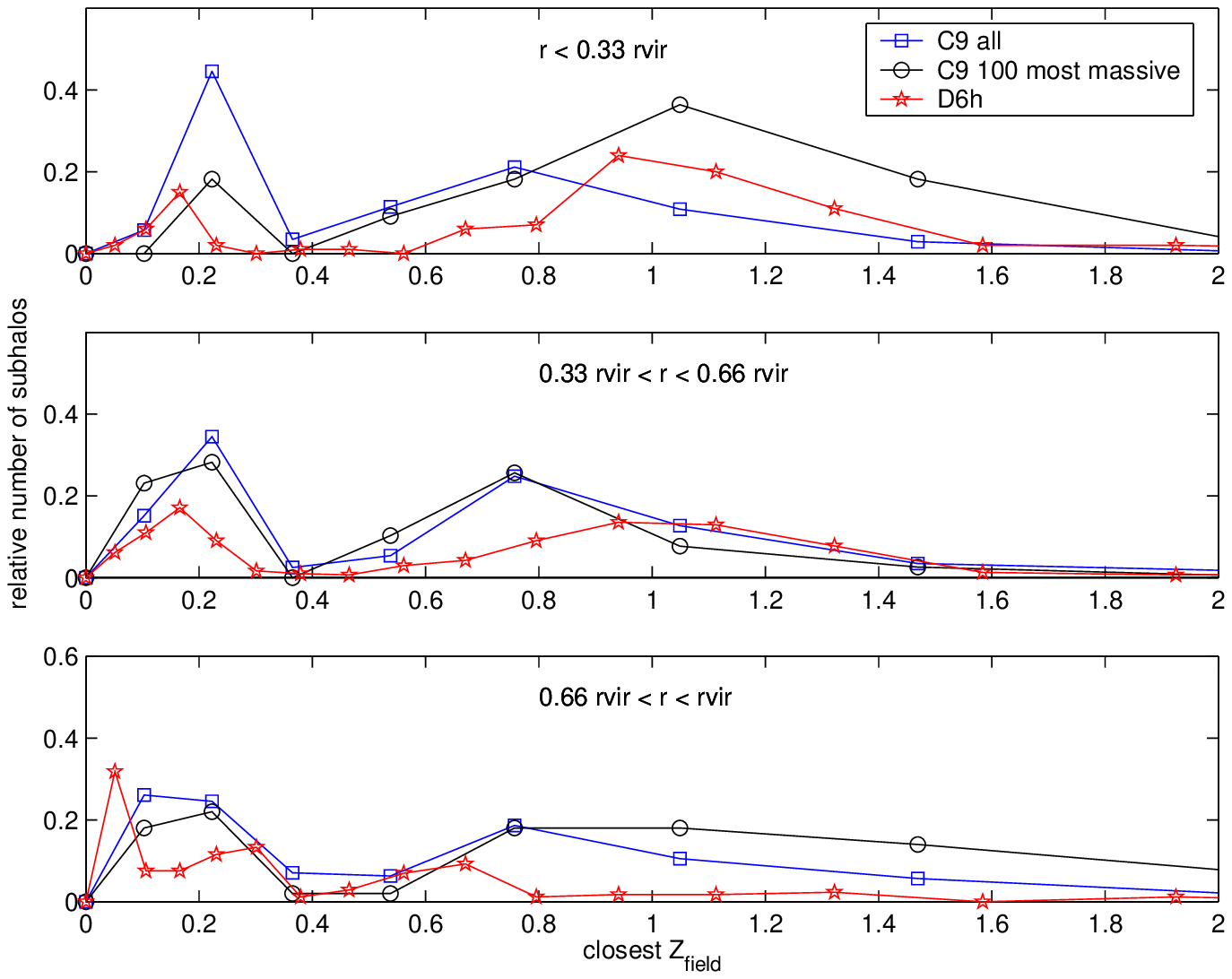}
\caption{
Histograms of accretion redshifts of subhalos in
two high resolution $\Lambda$CDM cluster simulations ($D6h$ and $C9$,
same data as in Figure \ref{zaccretedscatter}).
The top, middle and bottom panels correspond to the inner, intermediate
and outer regions of the two clusters.}
\label{zaccreted}
\end{figure}

>From the scatter plot and also from the histogram of accretion redshifts in three
radial bins (Figure \ref{zaccreted}) one can see that the accretion rate is 
not a simple function of time
but there are epochs of very rapid or of very slow accretion. Both clusters show 
very little accretion around redshift 0.4, which seems to be a coincidence. 

The inner subhalos were accreted slightly earlier on average, in run D6h
the mean and standard deviation of expansion factors at accretion 
is $a=0.59\pm0.14$ for subhalos that end up in the inner 33 percent of the 
cluster and $a=0.80\pm0.16$ for the outer 33 percent. For run C9 however all three 
radial bins give a mean of about $a=0.7$. More halos must be analyzed to see
if there really is a correlation of accretion redshift with cluster-centric radius,
but we can already say that such a correlation must be weak and have
a very large scatter. 

\section{Pericenters of halos in the outskirts of clusters}

How many galaxies in the outskirts of clusters have passed trough the inner, hot dense
part of the cluster and how many are approaching the system for the first time? 
This question is interesting since some spiral galaxies in the outskirts of the Virgo cluster 
are observed to be deficient in neutral Hydrogen. First attempts to answer this
questions include tracing back {\it particles} in cosmological Nbody simulations 
(\cite[Balogh  \etal\ 2000]{Balogh}) and analytical, spherical infall and rebound calculations 
(\cite[Mamon \etal\ 2004]{Mamon}).   

We traced back all subhalos and halos around the cluster D6h and 
measured the distance to the cluster core going back to the formation epoch of 
the cluster (z$\simeq$0.6). The interval with a time resolution of 0.6 Gyrs.

\begin{figure}
\includegraphics{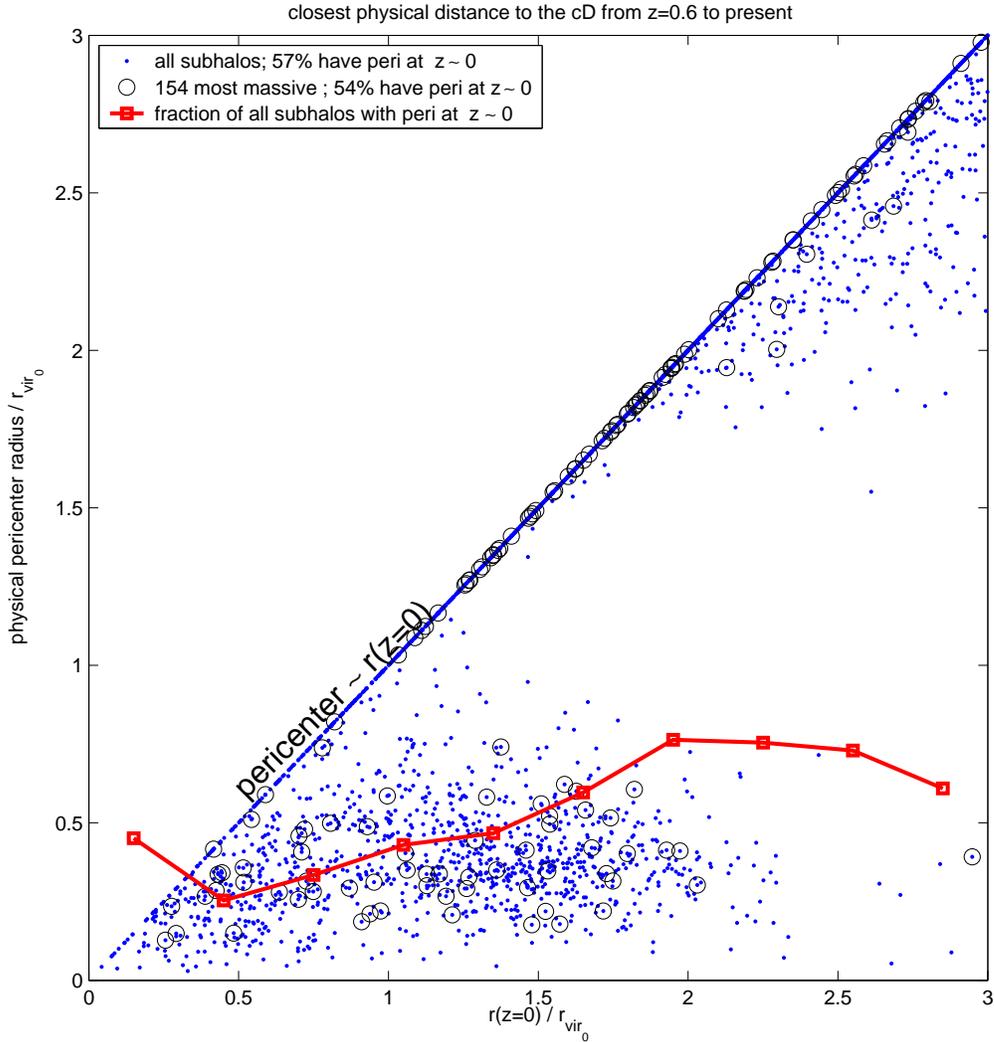}
\caption{Closest physical distance to the cluster center form z=0.6 to present 
vs. distance form the cluster center today. The red line gives the fraction of
(sub)halos that have their pericenter today (within the time resolution of
0.6 Gyrs). The points in the lower right corner show that many halos well
outside the cluster today passed deep through it earlier.
}\label{rperiscatterHC}
\end{figure}

Figure \ref{rperiscatterHC}  shows the pericenter distance of the (sub)halos versus 
cluster-centric distance today. The points on the diagonal are halos that have their
pericenter at z$\simeq$0, the halos just below the diagonal in the upper right corner
of the Figure are orbiting two satellite groups a distances of about 2 $r_{\rm virial}$.
In the lower left corner ($<r_{\rm virial}$) we see todays subhalos.
For ($r>r_{\rm virial}$) there is a large 
population of halos that have pericenters well within the
cluster. These are halos in the outskirts of the cluster which have passed through the
cluster earlier. About half of the halos between $r_{\rm virial}$ and 2 $r_{\rm virial}$
have a pericenter smaller than $r_{\rm virial}$. Most of them (at least 70 percent
\footnote{Note that in the inner part of the cluster the dynamical times become 
comparable to the interval between outputs, so the real pericenters will be smaller.})
have even passed through the inner part of the cluster ($r < 0.5 r_{\rm virial}$).
Finally the points in the lower right part of the plot show that in some rare cases 
halos that passed through the cluster can rebound out to 3 $r_{\rm virial}$, which
is a little larger than the maximal distance of 2.5 $r_{\rm virial}$ obtained
from analytical, spherical infall and rebound calculations (\cite{Mamon}). 

\section{Spatial distribution of subhalos}

\cite{Ghigna} showed that the spatial distribution of subhalos is antibiased
with respect to the mass. \cite[Diemand \etal\ (2004a)]{Diemand2004a} 
confirmed that this was not
a resolution effect but most likely due to physical overmerging of dark 
matter halos as they entered the central cluster region. In order to reproduce
the observed spatial distribution of galaxies (see Figure 5), 
dissipation is likely to
play a key role (\cite[Gao \etal\ 2004b]{Gao2004b}). 
It is not expected that dissipation will greatly alter
the internal structure of galactic halos hosting disks with type later than Sb. 
These galaxies will suffer the same fate as the infalling subhalos and become tidally
disrupted by the cluster environment. This immediately leads to the 
morphology-density/radius relation since only ellipticals and Sa/Sb galaxies can
survive near the cluster centre. Ellipticals will be especially dense since the multiple
merging of gas rich proto-galaxies will undoubtedly lead to strong gas inflow 
into the central regions. The inter-galactic medium 
will rapidly be stripped from the central
Sa/Sb galaxies and combined with a moderate amount of disk heating from tides these
galaxies will rapidly turn into S0's. However it is hard to distinguish this
scenario from one which disk formation is suppressed within the proto-cluster
environment.

\begin{figure}
\includegraphics{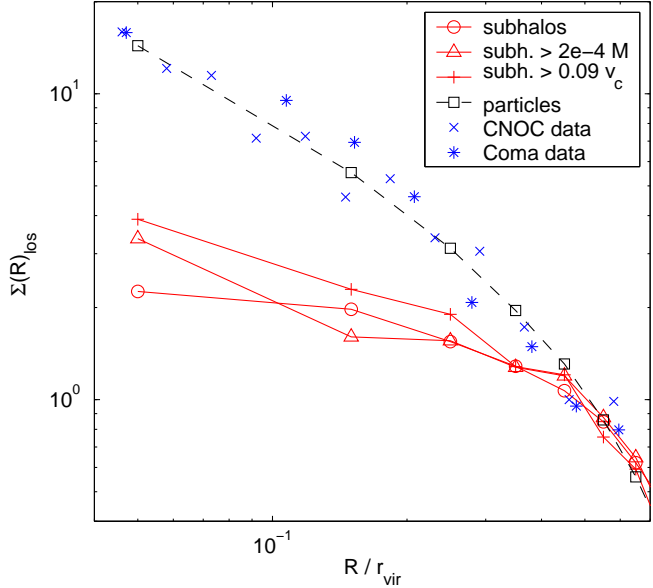}
\caption{
The distribution of subhalos plotted against the smooth dark matter component.
The symbols show the radial distribution of galaxies from a sample of CNOC
clusters (\protect\cite{Carlberg}) and the Coma cluster 
(\protect\cite{Lokas}).
}\label{spatial}
\end{figure}

\section{Conclusions}\label{sec:concl}

Numerical simulations that follow only the dark matter component have provided
numerous insights into the dynamical evolution of substructures. From these
simulations we can infer a great deal about the environmental processes that
may have affected galaxies both within the cluster and in the surrounding regions.
However it is also clear that dissipation must play an important role in enabling galaxies
to survive in the harsh environment within the inner regions of the virialised cluster.
We summarise some of the results of our CDM cluster simulations here and look foward
to enormous progress over the next decade in hydro-dynamical simulation of
galaxy formation in different environments.

\begin{itemize}
\item The average accretion redshift of subhalos does not change significantly with
the final time cluster-centric distance, i.e. there is no strong age-radius correlation
in $\Lambda$CDM {\it subhalos} (however this does not exclude an
age-radius correlation for {\it cluster galaxies}, since they do not 
trace the subhalos in a simple one-to-one correspondence: 
the subhalo number density profile are much shallower).
\item About 50 percent of the galaxies that have a distance between one and two
virial radii form the cluster center today have passed trough the cluster earlier.
\item Most of them (at least 70 percent) even approached the cluster center to less
than half of the virial radius.
\item There are some (rare) cases where a halo passes trough the inner part of the cluster
and then rebounds out to three virial radii. 
\item Dissipation must play an important role in enabling galaxies to survive
in the central cluster regions. The morphology-density relation may be due to
the disruption of disks at the cluster centre.
\end{itemize}

\begin{acknowledgments}
J.D. is supported by the Swiss National Science Foundation. Simulations were carried out
on the zBox supercomputer.
\end{acknowledgments}

\end{document}